\documentclass[a4paper,superscriptaddress,twocolumn,prx]{revtex4}

\usepackage[pdftex]{graphicx}

\begin{document}

\title{A quantum access network}

\author{Bernd Fr\"ohlich}
%\email{thiago.schiavo@gmail.com}
%\homepage{http://legauss.blogspot.com}
\affiliation{Toshiba Research Europe Ltd, 208 Cambridge Science Park, Cambridge CB4 0GZ, UK}
\affiliation{Corporate Research \& Development Center, Toshiba Corporation, 1 Komukai-Toshiba-Cho, Saiwai-ku, Kawasaki 212-8582, Japan}

\author{James F Dynes}
%\email{thiago.schiavo@gmail.com}
%\homepage{http://legauss.blogspot.com}
\affiliation{Toshiba Research Europe Ltd, 208 Cambridge Science Park, Cambridge CB4 0GZ, UK}
\affiliation{Corporate Research \& Development Center, Toshiba Corporation, 1 Komukai-Toshiba-Cho, Saiwai-ku, Kawasaki 212-8582, Japan}

\author{Marco Lucamarini}
%\email{thiago.schiavo@gmail.com}
%\homepage{http://legauss.blogspot.com}
\affiliation{Toshiba Research Europe Ltd, 208 Cambridge Science Park, Cambridge CB4 0GZ, UK}
\affiliation{Corporate Research \& Development Center, Toshiba Corporation, 1 Komukai-Toshiba-Cho, Saiwai-ku, Kawasaki 212-8582, Japan}

\author{Andrew W Sharpe}
%\email{thiago.schiavo@gmail.com}
%\homepage{http://legauss.blogspot.com}
\affiliation{Toshiba Research Europe Ltd, 208 Cambridge Science Park, Cambridge CB4 0GZ, UK}

\author{Zhiliang L Yuan}
%\email{thiago.schiavo@gmail.com}
%\homepage{http://legauss.blogspot.com}
\affiliation{Toshiba Research Europe Ltd, 208 Cambridge Science Park, Cambridge CB4 0GZ, UK}
\affiliation{Corporate Research \& Development Center, Toshiba Corporation, 1 Komukai-Toshiba-Cho, Saiwai-ku, Kawasaki 212-8582, Japan}

\author{Andrew J Shields}
%\email{thiago.schiavo@gmail.com}
%\homepage{http://legauss.blogspot.com}
\affiliation{Toshiba Research Europe Ltd, 208 Cambridge Science Park, Cambridge CB4 0GZ, UK}
\affiliation{Corporate Research \& Development Center, Toshiba Corporation, 1 Komukai-Toshiba-Cho, Saiwai-ku, Kawasaki 212-8582, Japan}

\begin{abstract}
The theoretically proven security of quantum key distribution (QKD) could revolutionise how information exchange is protected in the future\cite{Lutkenhaus2009,Scarani2009}. Several field tests of QKD have proven it to be a reliable technology for cryptographic key exchange and have demonstrated nodal networks of point-to-point links\cite{Elliott2005,Peev2009,Sasaki2011}. However, so far no convincing answer has been given to the question of how to extend the scope of QKD beyond niche applications in dedicated high security networks. Here we show that adopting simple and cost-effective telecommunication technologies to form a \textit{quantum access network} can greatly expand the number of users in quantum networks and therefore vastly broaden their appeal. We are able to demonstrate that a high-speed single-photon detector positioned at a network node can be shared between up to 64 users for exchanging secret keys with the node, thereby significantly reducing the hardware requirements for each user added to the network. This point-to-multipoint architecture removes one of the main obstacles restricting the widespread application of QKD. It presents a viable method for realising multi-user QKD networks with resource efficiency and brings QKD closer to becoming the first widespread technology based on quantum physics.
\end{abstract}

%\keywords{latex-community, revtex4, aps, papers}

\maketitle

%% make sure you have the nature.cls and naturemag.bst files where
%% LaTeX can find them

%\bibliographystyle{naturemag}

%% Notice placement of commas and superscripts and use of &
%% in the author list

%\begin{document}

%\maketitle

%\begin{affiliations}
% \item Toshiba Research Europe Ltd, 208 Cambridge Science Park, Cambridge CB4 0GZ, UK
% \item Corporate Research \& Development Center, 1 Komukai-Toshiba-Cho, Sawai-ku, Kawasaki 212-8582, Japan
%\end{affiliations}

In a nodal QKD network multiple trusted repeaters are connected via point-to-point links between a quantum transmitter (Alice) and a quantum receiver (Bob). These point-to-point links can be realised with long-distance optical fibres, and in the future might even utilize ground to satellite communication\cite{R.2007,Nauerth2013,Wang2013}. While point-to-point connections are suitable to form a backbone \textit{quantum core network} to bridge long distances, they are less suitable to provide the last-mile service needed to give a multitude of users access to this QKD infrastructure. Reconfigurable optical networks based on optical switches or wavelength-division multiplexing have been suggested to achieve more flexible network structures\cite{Elliott2005,Toliver2003,Chapuran2009,Chen2010,Wang2010}, however, they also require the installation of a full QKD system per user, which is prohibitively expensive for many applications.

Giving a multitude of users access to the nodal QKD network requires point-to-multipoint connections. In modern fibre-optic networks point-to-multipoint connections are often realized passively using components such as optical power splitters\cite{ITU2008}. Single photon QKD with the sender positioned at the network node and the receiver at the user premises\cite{Townsend1997} lends itself naturally to a passive multi-user network (see Fig.~1\textbf{a}). However, this \textit{downstream} implementation has two major shortcomings. Firstly, every user in the network requires a single photon detector, which are often expensive and difficult to operate. And secondly, it is not possible to deterministically address a user. All detectors therefore have to operate at the same speed as the transmitter in order not to miss photons, which means most of the detector bandwidth is unused.

\begin{figure} [htbp]
\centering
\includegraphics[width=\columnwidth]{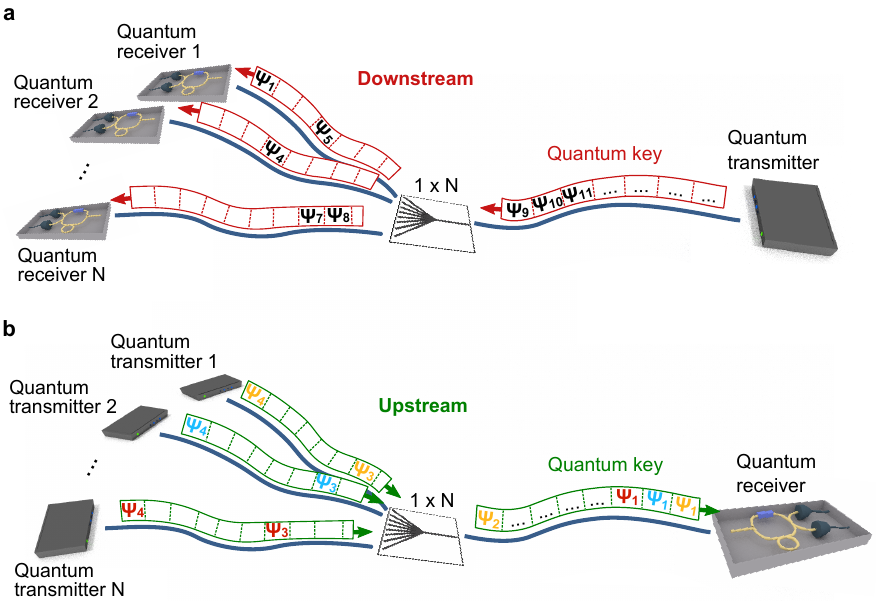}
\caption{Downstream and upstream quantum access network. \textbf{a} In a downstream configuration the quantum transmitter is positioned at the network node. The transmitted quantum key is randomly directed to one of the quantum receivers by a passive optical splitter. Each user needs a single-photon detector and the key is not distributed deterministically. \textbf{b} The upstream configuration requires only a single detector at the network node. The quantum transmitters share this detector by ensuring that only photons from one transmitter at a time reach the receiver.}
\label{fig:1}
\end{figure}

Here, we show that both problems associated with a \textit{downstream} implementation can be overcome with a conceptual advancement: the most valuable resource should be shared by all users and should operate at full capacity. We propose and demonstrate an \textit{upstream} quantum access network, in which the transmitters are placed at the end user location and a common receiver is placed at the network node as shown in Fig.~1\textbf{b}. A careful study of the cross-talk between senders arising from the shared receiver topology shows that operation with up to 64 users is feasible, which we demonstrate by performing multi-user QKD over a 1x64 passive optical splitter. The results presented here for the first time highlight a practical and viable approach to extend the scope of QKD applications to many more users. Our approach would also be advantageous in a fully quantum network in which a quantum relay or repeater is located at the common node.

One of the main challenges for realizing an \textit{upstream} quantum access network is to develop independently operating quantum transmitters which exchange secure keys efficiently with the receiver in parallel. For example, active stabilisation in QKD systems is typically implemented at the receiver side\cite{Dixon2010}. In our scheme, however, the receiver is a reference for multiple transmitters and phase changes hence have to be pre-compensated by each user individually. Figure~2 shows a schematic of our experimental setup (see also methods). We developed two flexible quantum transmitters which can operate at varying repetition rates and contain all the stabilisation components necessary for continuous operation. They also include additional polarisation control elements to achieve higher key rates (see methods). At the centre of the quantum network is a passive optical splitter which connects multiple transmitters to the receiver. The fibre distance between transmitters and receiver was chosen to be close to the maximum distance defined for gigabit passive optical networks in the International Telecommunication Union (ITU) standardization document\cite{ITU2008}. We use a phase-encoding BB84 QKD protocol with decoy states\cite{Hwang2003,Wang2005,Lo2005,Ma2005} implemented with asymmetric Mach-Zehnder interferometers and intensity modulators. The quantum receiver decodes the phase information with a matching interferometer and uses two high-speed detectors based on avalanche photo-diodes to detect the single photons with a rate of 1~GHz\cite{Yuan2007,Yuan2008}. We implement phase encoding as it is robust against fluctuations on the transmission channel and allows for a simple stabilisation mechanism.

\begin{figure*} [htbp]
\centering
\includegraphics[width=\textwidth]{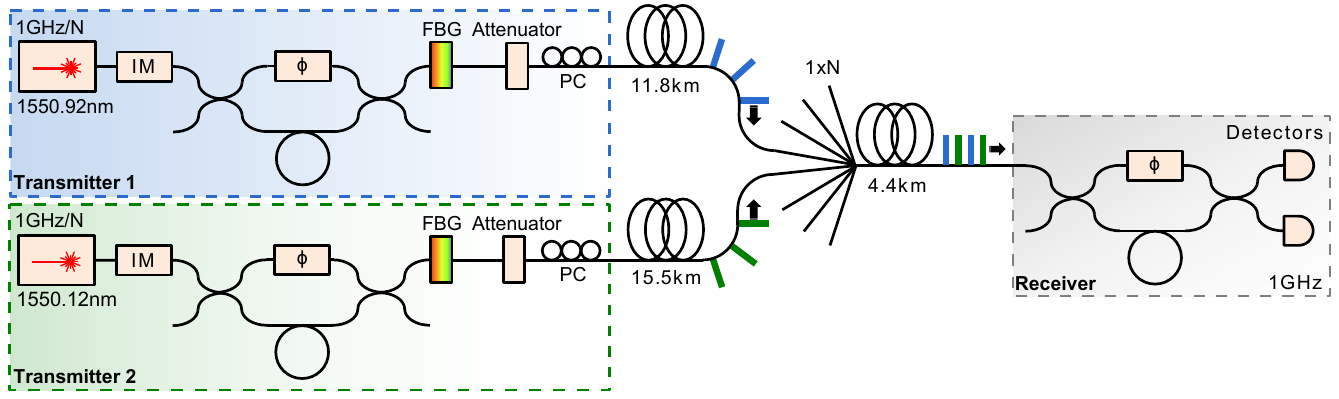}
\caption{Experimental setup. Two quantum transmitters are connected to a single quantum receiver via a passive optical splitter and fibre spools. Each transmitter encodes bit and basis information on short laser pulses with an asymmetric Mach-Zehnder interferometer. The intensity of the pulses is modulated with an intensity modulator (IM) and attenuated to the single photon level. A polarisation controller (PC) pre-compensates the polarisation and a fibre-Bragg-grating (FBG) compensates pulse broadening in the fibre. The receiver decodes the phase information with a matching interferometer. The two outputs of this interferometer are connected to single-photon detectors which operate at 1~GHz.}
\label{fig:2}
\end{figure*}

In order to share a single-photon detector between multiple transmitters we adopt a novel and efficient scheme which allows continuous and stable exchange of keys necessary to reduce the detrimental effect of finite size samples\cite{Hayashi2007,Scarani2008,Scarani2008a,Cai2009,Lucamarini2012}. Large fluctuations during a key session will reduce the amount of sifted bits that are transmitted and therefore reduce the amount of secure bits that can be distilled after privacy amplification. In our scheme all quantum transmitters operate continuously in parallel allowing for uninterrupted key sessions. We operate each transmitter at a fraction of the speed of the receiver, for example 1~GHz/8=125~MHz in an 8 user network. The transmitters are synchronised such that their pulses fall into subsequent detection time slots and can be clearly assigned to each user as shown in Fig.~1\textbf{b}. This scheme has two main advantages: Firstly, polarization, phase, and synchronisation tracking is done continuously by each QKD transmitter against the common quantum receiver, thus allowing stable operation of the quantum network. And secondly, the transmitter can be realised with simpler electronics and optics due to the lower operational speed.

\begin{figure} [htbp]
\centering
\includegraphics[width=\columnwidth]{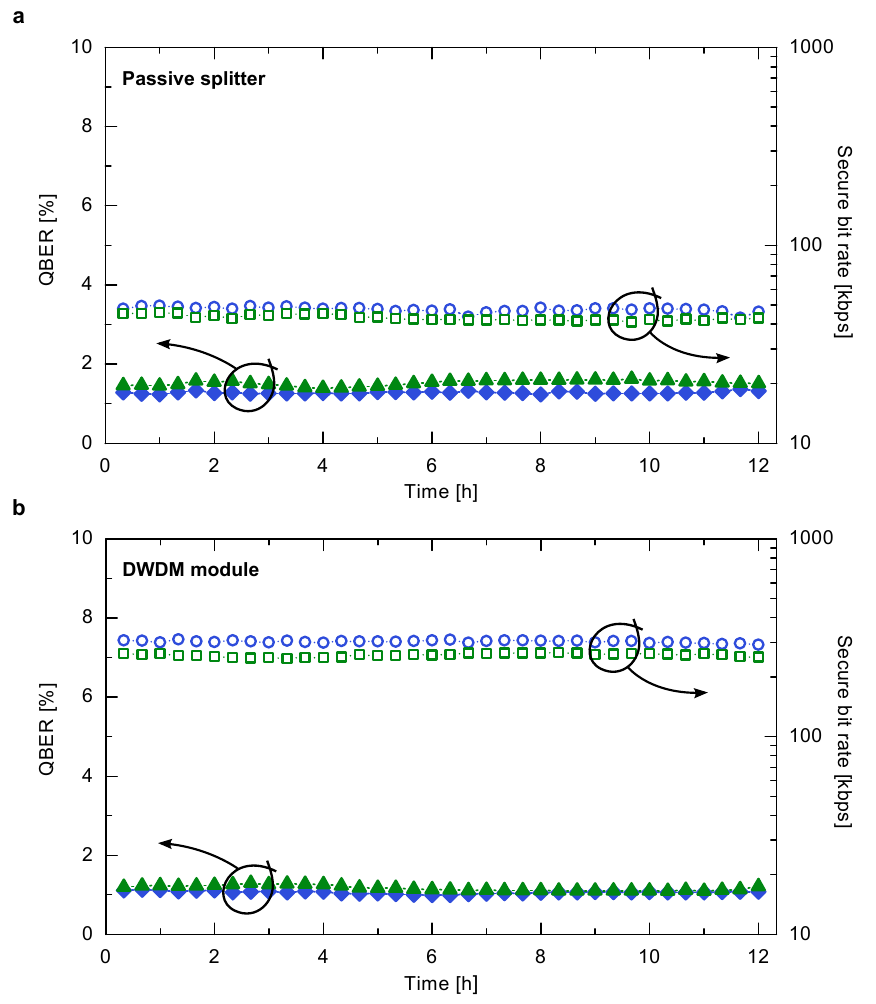}
\caption{Stable operation of quantum access network. \textbf{a} Quantum bit error ratio (QBER) (transmitter 1: filled blue diamonds, transmitter 2: filled green triangles) and secure bit rate (transmitter 1: open blue circles, transmitter 2: open green squares) for each 20~minute key session in a network supporting 8 users with a 1x8 passive optical splitter. \textbf{b}  QBER and secure bit rate for each 20~minute key session in an 8 user network using dense wavelength-division multiplexing (DWDM) optics.}
\label{fig:3}
\end{figure}

In a first experiment we demonstrate stable operation over 12~hours of a 1x8 quantum access network which is populated by 2 users. We operate both transmitters at 125~MHz and use a passive 1x8 splitter to combine their signals, the total transmission loss including the splitter is 13.6~dB (14~dB) for transmitter 1 (transmitter 2). Counts in each detection gate are allocated either to one of the transmitters or identified as an empty gate depending on their timing information. In a 20~minute key session we record almost 300~Mbit of counts per transmitter. Figure~3\textbf{a} shows the quantum bit error ratio (QBERs) and secure bit rate for each key session. Although assigned to a specific transmitter, both the transmitter and the receiver subsystem contribute to the QBER. Due to the low QBERs (transmitter 1: 1.28~\%, transmitter 2: 1.53~\%) we can exchange secure bits very efficiently with an average secure bit rate of 47.5~kbps (43.1~kbps) for transmitter 1 (transmitter 2). Continuous operation over a month would allow unconditionally secure one-time-pad encryption of more than 10~GByte of data for each user, which is enough for example to protect over one hundred thousand emails.

The key rate can be increased further by the use of wavelength-division multiplexing optics instead of passive splitters because of the lower insertion loss of these devices. We demonstrate this in a second experiment by replacing the 1x8 splitter with an 8 channel thin-film dense-wavelength-division multiplexing module. The specified loss of the multiplexing module of $\approx2.5$~dB is 5 times lower than for the 1x8 splitter and accordingly leads to a proportional increase of the count rate. In this experiment the quantum transmitters do not need to be modified as we designed the emission wavelengths of transmitter 1 and transmitter 2 to coincide with channels 33 ($1550.92$~nm) and 34 ($1550.12$~nm) of the ITU grid, respectively. Figure~3\textbf{b} shows the resulting QBER and secure bit rate for each key session. In addition to the higher transmission rate, the ratio of dark counts to photon counts decreases leading to a reduction of the QBER (transmitter 1: 1.06~\%, transmitter 2: 1.17~\%) and therefore to an even higher increase of the secure bit rate (6-fold), corresponding to almost 100~GByte of key material per month (transmitter 1: 303~kbps, transmitter 2: 259~kbps). 

We can extrapolate the performance of a network with more users by studying the cross-talk between two transmitters in detail. To determine the cross-talk we measure the average count rate of one transmitter with the other transmitter either on or off $C_{\rm on}$ and $C_{\rm off}$, respectively. As shown in the inset of Fig.~4\textbf{a} the transmitters are operated at 125~MHz allowing us to vary the gate separation between them. From this data we extract how many spurious detection events the second transmitter (green) causes in the detection gates allocated to the first transmitter (blue). Figure~4\textbf{a} displays the relative count rate increase $(C_{\rm on}-C_{\rm off})/C_{\rm off}$ for various gate separations of transmitter 1 and 2. There are two effects contributing to the increase: Firstly, detection of pulses sent from the second transmitter causes afterpulses in the gate periods of the first transmitter. This leads to an increase of the count rate per added user independent of the gate separation by $p_{\rm A}/N$ as indicated by the dashed line, where $p_A$ is the afterpulse probability of the detector and $N$ is the number of users the network supports. Secondly, smaller gate separation between transmitter 1 and 2 increases the cross-talk which is most likely due to late arrival of photons or ringing of the detector electronics. 

\begin{figure*} [htbp]
\centering
\includegraphics[width=\textwidth]{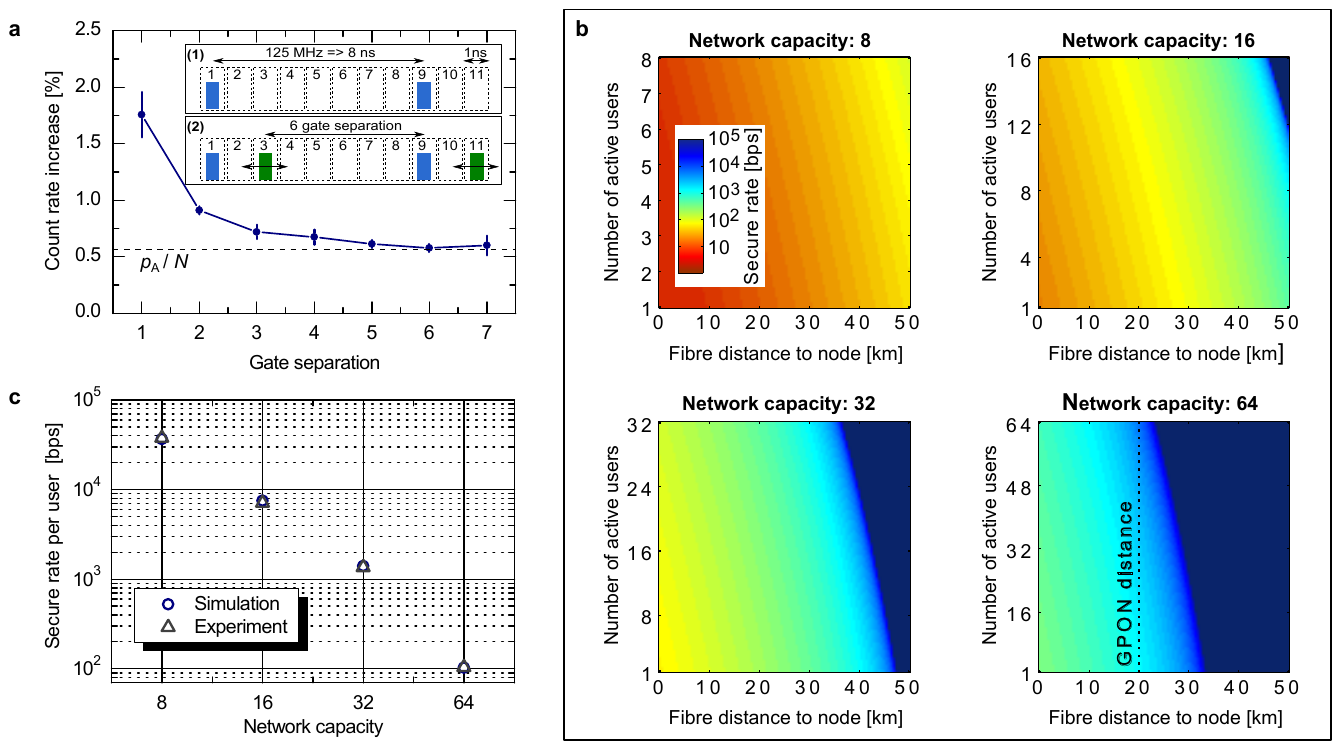}
\caption{Quantum access network with varying capacity. \textbf{a} Relative count rate increase due to cross-talk between the transmitters as a function of gate separation. The dashed line indicates the limit given by detector afterpulses. Error bars correspond to one standard deviation of three consecutive measurements. Inset: Measurement principle (see text). \textbf{b} Simulation of the secure bit rate per user as a function of fibre distance and number of active users in the network for various network capacities (see text and methods). \textbf{c} Secure bit rate per user in a quantum network with varying capacity estimated from a two user measurement at 500~MHz.}
\label{fig:4}
\end{figure*}

Using the cross-talk data we simulate how the key rate of a single transmitter changes when more users are added to the network (see also methods). Figure~4\textbf{b} shows logarithmic colour-scale plots of the secure bit rate per user as a function of fibre distance from the transmitters to the node and number of active users in the network. We simulate the rate for various network capacities which are given by the splitting ratio of the passive optical splitter installed in the system. For the simulation we assume that each active transmitter operates with 1~GHz/N and that the fibre distance to the node is the same for all transmitters. The data shows clearly that even for a 64 user network, which inherently has a loss of approximately 20~dB from the 1x64 splitter, secure transmission is possible up to the maximum distance for gigabit passive optical networks (GPON) of 20~km with all users active. In networks supporting fewer users there is a margin to allow longer fibre distances or, correspondingly, higher loss in the system.

To verify this result experimentally we switch to 500~MHz operation of the transmitters and vary the splitting ratio from 1x8 to 1x64. Operation at 500~MHz with two transmitters allows us to emulate a fully occupied network as photon detections are possible in all detector time slots. Figure~4\textbf{c} displays the estimated key rate per user based on the measured secure bit rates of transmitter 1 and 2 and how it compares to the expected value from the simulation. For the key rate estimation we add the key rate of the two transmitters and divide it by the network capacity. The data confirms the result obtained from the simulation as it demonstrates that a 64 user network is feasible with our scheme.

In conclusion, we have demonstrated that passive optical networks have the potential to scale up the number of users in a nodal QKD network. We have shown that the reduction of the secure key rate accompanying a time-division multiplexing approach can be greatly mitigated by using a high speed single-photon detector. The network node in our scheme acts as a receiver and has to be trusted intrinsically, however, techniques such as classical secret sharing\cite{Barnett2011a} or measurement-device-independent QKD\cite{Lo2012} might be used to relax this requirement in the future. It might also be possible to combine classical data transport on the same fibre in quantum-secured access networks\cite{Patel2012}. Quantum access networks could initially find application for example to protect smart community or smart grid networks allowing authenticated data collection from multiple locations in a critical infrastructure network\cite{Hughes2013}.

\section{methods}

\subsection{Experimental setup.}
Each quantum transmitter consists of a source of short laser pulses, an intensity modulator (IM), an asymmetric Mach-Zehnder interferometer including a phase modulator in one arm, a fibre Bragg grating (FBG), an attenuator and a polarisation controller (see Fig.~2). The laser source is a distributed feedback laser generating laser pulses shorter than 50~ps with a selectable repetition rate of up to 1~GHz. The wavelengths of transmitter 1 and transmitter 2 are tuned to coincide with channels 33 and 34, respectively, of the grid defined by the International Telecommunication Union (ITU). The intensity modulator in combination with the attenuator at the output sets the power of each pulse to one of three power levels necessary for the decoy protocol$^{16-19}$: 0.5 photons/pulse for signal pulses, 0.1 photons/pulse for decoy pulses, and 0.0002 photons/pulse for vacuum pulses. We send signal, decoy, and vacuum pulses with probabilities of 98.83~\%, 0.78~\%, and 0.39~\%, respectively.

The asymmetric Mach-Zehnder interferometer is made of off-the-shelf fibre optic components and the length difference between long and short arms is matched to the receiver interferometer with a tunable optical delay. The components do not require temperature stabilisation or vibration isolation. Drifts of the relative phase between long and short arm of the interferometer with respect to the receiver interferometer are compensated by applying a DC bias across the phase modulator. Each transmitter compensates the phase difference individually and independently of other users in the network. No compensation is necessary in the receiver interferometer, which acts as a phase reference. The output beam splitter of the transmitter interferometer and the input beam splitter of the receiver interferometer are polarising to direct the photons into the correct arm of the receiver interferometer, thus avoiding a 3~dB penalty when using polarisation insensitive interferometers. We therefore additionally pre-compensate the polarisation of the transmitted pulses with a polarisation controller (PC) to achieve maximum count rate at the receiver. The fibre Bragg grating compensates pulse broadening across the fibre link to avoid a further 1.5~dB decrease of the count rate.

The quantum receiver consists of a reference Mach-Zehnder interferometer which decodes the phase information of the pulses sent from the transmitters and detects the photons with InGaAs single-photon avalanche photodiodes (APDs) with a rate of 1~GHz using a self-differencing technique$^{20,21}$. The APDs are operated at a temperature of~$-30$~$\deg$~C. A typical detection efficiency is 15~\%, with $8\times 10^{-6}$ dark counts per gate, and an afterpulse probability of 4.5~\%. Detection events are sorted into $N$ time bins, where $N$ is the number of users the network supports. Each transmitter is aligned relative to a master clock from Bob such that their photons coincide with a specific time bin and can be clearly assigned to a user. A variable time delay implemented in each transmitter allows to control the alignment continuously with a feedback signal generated from the detector count rate.

\subsection{Secure key rate.}

We implement the standard BB84 protocol with decoy states in our setup. The quantum transmitter prepares one of four phase states 0, $\frac{\pi}{2}$, $\pi$, and $\frac{3\pi}{2}$ with equal probabilities and the receiver chooses either phase 0 or $\frac{\pi}{2}$. All events with non-matching basis are discarded in the sifting process. Based on the individual error rates of signal, decoy, and vacuum states we estimate single-photon parameters for each user individually. Our security analysis$^{26}$ takes finite-size effects into account$^{23-25}$ and achieves bit rates close to the asymptotic limit for key sessions of 20~min.  The secure key rate is lower bounded by the following quantity
\begin{equation}
R = \{Q_1[1-H(e_1)]-Q f_{\rm EC}(e) H(e)+Q_0-\Delta\}/t \,.
\end{equation}
Here $Q_1$ is the estimated number of sifted bits from single-photon states, $H(e_1)$ the binary entropy function of the estimated error rate of those bits, $Q$ the total number of sifted bits, $f_{\rm EC}$ the error correction efficiency which is set to 1.1, $e$ is the QBER of sifted
bits, $Q_0$ is the estimated number of sifted bits originating from vacuum pulses, and $t$ is the key session time. Finite size effects are included by subtracting $\Delta$ which is proportional to $\sqrt{Q}$ and to $\log_2(\epsilon^{-1})$, where $\epsilon$, equal to $10^{-10}$ in our system, is related to the overall security of the system$^{26}$. 

Parameters of the protocol such as decoy level and decoy probability have to be chosen carefully to achieve optimal secure key rates. For example, the estimation of $Q_1$ depends directly on the chosen decoy photon flux and probability. We simulate the achievable secure key rates in advance to select suitable parameters for the experiment. For convenience we use one set of parameters for all measurements presented here which we found to lead to stable results in all considered configurations.

\subsection{Simulation.}
We simulate the secure key rate per user in a network which is populated by more than two users based on measured experimental parameters. For the simulation we assume that each active transmitter operates with 1~GHz/N and that the fibre distance to the node is the same for all transmitters. Starting point is calculating the probability to get an error count for a single transmitter in an otherwise empty network using
\begin{equation}
\eta_{\rm err} = \eta\left(e_{\rm opt}+\frac{p_{\rm A}}{2N}\right)+\frac{p_{\rm D}}{2}\,.
\end{equation}
Here, $e_{\rm opt}$ is the optical error due to encoding imperfections, $p_{\rm A}$ is the afterpulse probability and $p_{\rm D}$ the dark count probability of the detector, and the detection probability $\eta$ is given by $\eta = \mu 10^{-0.2L/10}l_{\rm spl} \eta_{\rm Bob}$, with $\mu$ the photon flux, $L$ the total fibre length in kilometers, $l_{\rm spl}$ the splitter loss, and $\eta_{\rm Bob}$ the detection efficiency of the receiver. We use the following parameters in the simulation: $e_{\rm opt}=0.5$~\%, $p_{\rm A}=4.5$~\%, $p_{\rm D}=2\cdot 8\times 10^{-6}$, and $\eta_{\rm Bob}=9.04$~\%.

Adding more users to the network will increase the error rate due to cross-talk between the users. The QBER including cross-talk counts is then given by
\begin{equation}
e=\frac{\eta_{\rm err}+\frac{1}{2}p_{\rm X}(n-1)\eta}{\eta\left(1+\frac{p_{\rm A}}{N}\right)+p_{\rm D}+p_{\rm X}(n-1)\eta}
\end{equation}
with $n$ the number of active users in the network. The average count rate increase per added user
\begin{equation}
p_{\rm X}=\frac{1.9\%}{N-1}+\frac{p_{\rm A}}{N} \,.
\end{equation}
is extracted from the data shown in Fig.~4\textbf{a} by taking both the base increase of $p_{\rm A}/N$ and the enhanced cross-talk at short gate separation into account. For the splitter loss $l_{\rm spl}$ we use our measured values of 9.7~dB, 13~dB, 16.1~dB, and 19.5~dB for 1x8, 1x16, 1x32, and 1x64 splitting ratio, respectively.

The secure key rate is determined from a refined analysis calculating the QBER for all three signal levels of the decoy protocol. We use the same routine as for the experimental data to determine the secure key rate from these values. For the 32 and 64 user network we increased the key session length from 20~min to 2~h and 12~h, respectively, to compensate for the decreasing sample size due to the slower operation of the transmitter. Our measurements shown in Fig.~3 clearly indicate that longer key sessions are feasible. For the comparison of simulation and experiment shown in Fig.~4\textbf{c} we adopt the simulation to account for the non-equal fibre distances of the two transmitters, as well as the higher operational speed leading to different sample sizes compared to Fig.~4\textbf{b}.
%\end{methods}

%% Put the bibliography here, most people will use BiBTeX in
%% which case the environment below should be replaced with
%% the \bibliography{} command.

\bibliography{QANPaper}{}

%% Here is the endmatter stuff: Supplementary Info, etc.
%% Use \item's to separate, default label is "Acknowledgements"

%\begin{addendum}
% \item Put acknowledgements here.
% \item[Competing Interests] The authors declare that they have no
%competing financial interests.
% \item[Correspondence] Correspondence and requests for materials
%should be addressed to A.B.C.~(email: myaddress@nowhere.edu).
%\end{addendum}

%%
%% TABLES
%%
%% If there are any tables, put them here.
%%

\end{document}